\documentclass[aoas]{imsart}

\RequirePackage{amsthm,amsmath,amsfonts,amssymb,booktabs,url}
\RequirePackage[authoryear]{natbib}
\RequirePackage[colorlinks=True,
linkcolor=black,
filecolor=black,
urlcolor=black,
citecolor=black]{hyperref}
\RequirePackage{graphicx}
\RequirePackage{xurl}

\startlocaldefs

\endlocaldefs

\begin{document}

\begin{frontmatter}
\title{Probabilistic Inversion Modeling of Gas Emissions: A: Gradient-Based MCMC Estimation of Gaussian Plume Parameters}

\begin{aug}
\author[A]{\fnms{Thomas}~\snm{Newman}\ead[label=e1]{t.newman1@lancaster.ac.uk}\orcid{0009-0003-8453-9184}},
\author[A]{\fnms{Christopher}~\snm{Nemeth}\ead[label=e2]{c.nemeth@lancaster.ac.uk}\orcid{0000-0002-9084-3866}},
\author[B]{\fnms{Matthew}~\snm{Jones}\ead[label=e3]{matthew.m.jones2@shell.com}}
\and
\author[A,C]{\fnms{Philip}~\snm{Jonathan}\ead[label=e4]{p.jonathan@lancaster.ac.uk}\orcid{0000-0001-7651-9181}}


\address[A]{School of Mathematical Sciences, Lancaster University \printead[presep={ ,\ }]{e1,e2,e4}}
\address[B]{Shell Information Technology International B.V.\printead[presep={,\ }]{e3}}
\address[C]{Shell Information Technology International Ltd.\printead[presep={,\ }]{e4}}
\end{aug}

\begin{abstract}
 In response to global concerns regarding air quality and the environmental impact of greenhouse gas emissions, detecting and quantifying sources of emissions has become critical. To understand this impact and target mitigations effectively, methods for accurate quantification of greenhouse gas emissions are required. In this paper, we focus on the inversion of concentration measurements to estimate source location and emission rate. In practice, such methods often rely on atmospheric stability class-based Gaussian plume dispersion models. However, incorrectly identifying the atmospheric stability class can lead to significant bias in estimates of source characteristics. We present a robust approach that reduces this bias by jointly estimating the horizontal and vertical dispersion parameters of the Gaussian plume model, together with source location and emission rate, atmospheric background concentration, and sensor measurement error variance. Uncertainty in parameter estimation is quantified through probabilistic inversion using gradient-based MCMC methods. A simulation study is performed to assess the inversion methodology. We then focus on inference for the published Chilbolton dataset which contains controlled methane releases and demonstrates the practical benefits of estimating dispersion parameters in source inversion problems. 
\end{abstract}

\begin{keyword}
\kwd{Methane emissions}
\kwd{Bayesian inversion modeling}
\kwd{Manifold Metropolis-adjusted Langevin algorithm within Gibbs}
\kwd{Dispersion parameters}
\kwd{JAX}
\end{keyword}

\end{frontmatter}


\section{Introduction}\label{introduction}

The latest Intergovernmental Panel on Climate Change (IPCC) report concluded with high confidence that climate change is responsible for substantial damage to our ecosystems. We are approaching irreversible losses and can say with very high confidence that mass mortality events are being observed on land and in the oceans \citep{HLeeandJRomeroo}; Section 2.1.2 paragraph 3. Methane (CH4) has a global warming potential 84 times greater than carbon dioxide (CO2) over 20 years, making it a more powerful greenhouse gas \citep{IPCC2013climate}. With over 60\% of methane emissions being anthropogenic \citep{saunois2019global}, the current global average atmospheric CH4 concentration is about 1.93 parts per million (PPM) \citep{Lan2024Noaa}, increasing by over 0.075 PPM every decade \citep{nisbet2019very}. While carbon dioxide remains in the atmosphere for hundreds of years, methane's shorter atmospheric lifetime of 8.9 ± 0.6 years \citep{prinn1995atmospheric} (before it is chemically transformed or deposited out of the atmosphere to the earth's surface) means that reducing methane emissions can quickly mitigate global warming, aligning with the 2015 Paris Agreement's climate goals.

\vspace{2mm}

\noindent
Identifying and quantifying methane emissions leads to a better overview of sources of methane (e.g. leaks) which can subsequently be repaired or avoided. This plays a role in addressing climate change and enhancing sustainability efforts worldwide.  There is therefore a need to accurately estimate and report methane emissions from various sources, such as production facilities and distribution networks, where emissions may result from venting, flaring, or equipment leaks. For instance, through satellite observations, the Caspian coast of Turkmenistan has been identified as one of the most significant methane hotspots globally, a finding that has since received extensive media coverage. \cite{irakulis2022satellites} have linked these emissions to venting and pipeline leaks in oil and gas fields, presenting an opportunity to make informed decisions towards reducing CH4 emissions in the region.

\vspace{2mm}

\noindent
The Oil and Gas Methane Partnership 2.0 (OGMP 2.0) is the United Nations Environment Programme's voluntary framework for methane reporting and mitigation \citep{OGMP2024}. The program establishes 5 reporting levels of increasing granularity. To achieve gold standard reporting, operators must demonstrate their efforts to move towards level 5, which requires bottom-up emissions estimates from level 4 (based on an emission inventory combined with source-level measurements) to be reconciled with site-level measurements. Continuous monitoring of oil and gas facilities using methane concentration sensors is one method for obtaining site-level measurements, such measurements should be coupled with a robust inversion methodology in order to obtain accurate and trustworthy site emissions estimates. For many applications, including those to leak detection, it is reasonable to assume that the probability of a source existing in the domain under observation is small. Hence, inversion assuming at most a single source is likely to be adequate.

\vspace{2mm}

\noindent
Many inversion methods have been proposed to estimate source emission rates and locations, these can generally be grouped into optimization \citep{qiu2018atmospheric, albani2020source, wang2020hybrid} and Monte Carlo Markov Chain (MCMC) \citep{hirst2013locating, hirst2020methane, ma2021identifying, ijzermans2024long} algorithms. Forward models describing how gas disperses in the atmosphere can be used to attempt to explain measured gas concentrations. Inversion methods estimate the parameters that given the forward model would best describe the data collected. The most commonly used forward model for gas dispersion is the Gaussian plume model motivated by the solution of an underlying system of partial differential equations (PDEs) \citep{stockie2011mathematics}. The accuracy of the inversion is therefore closely linked to the accuracy of the forward model. The Gaussian plume model is very sensitive to the standard deviation of its Gaussian concentration distributions $\sigma_H$ and $\sigma_V$; we will now refer to these as ``wind sigma" parameters. In the literature, wind sigmas are often chosen based on the Pasquill atmospheric stability class (ASC) \citep{pasquill, cui2019investigating, mao2020impacts}. However, estimating the exact local ASC is often difficult in practice, and misspecifying it can substantially bias the inversion estimation. This paper shows that the bias introduced by ASC-based wind sigmas can be removed and source estimation improved by estimating wind sigmas instead of fixing them. \cite{mao2021improving} attempt this by optimizing the wind sigmas using a genetic algorithm over an ASC-based Briggs scheme parameterization. We propose to remove the dependence on ASCs and estimate wind sigmas within our MCMC procedure. A key challenge when accurately estimating source characteristics relies on adequately accounting for factors influencing the sensors' measurements. Background gas concentration and measurement error substantially influence the recorded concentration and can introduce bias in our estimation if not correctly accounted for.

\vspace{2mm}

\noindent
\textbf{\textit{Objectives:}} In this paper, we propose an MCMC inversion method jointly estimating source emission rates, locations, background concentrations, measurement error variance, and wind sigmas. We demonstrate that estimating the wind sigmas is beneficial for the practicing environmental modeler, in terms of improved inferences. Our methodology is based on the principles of probabilistic inversion, which allows us to incorporate uncertainties and prior knowledge effectively.

\vspace{2mm}

\begin{figure}
    \centering        \includegraphics[width=0.45\textwidth]{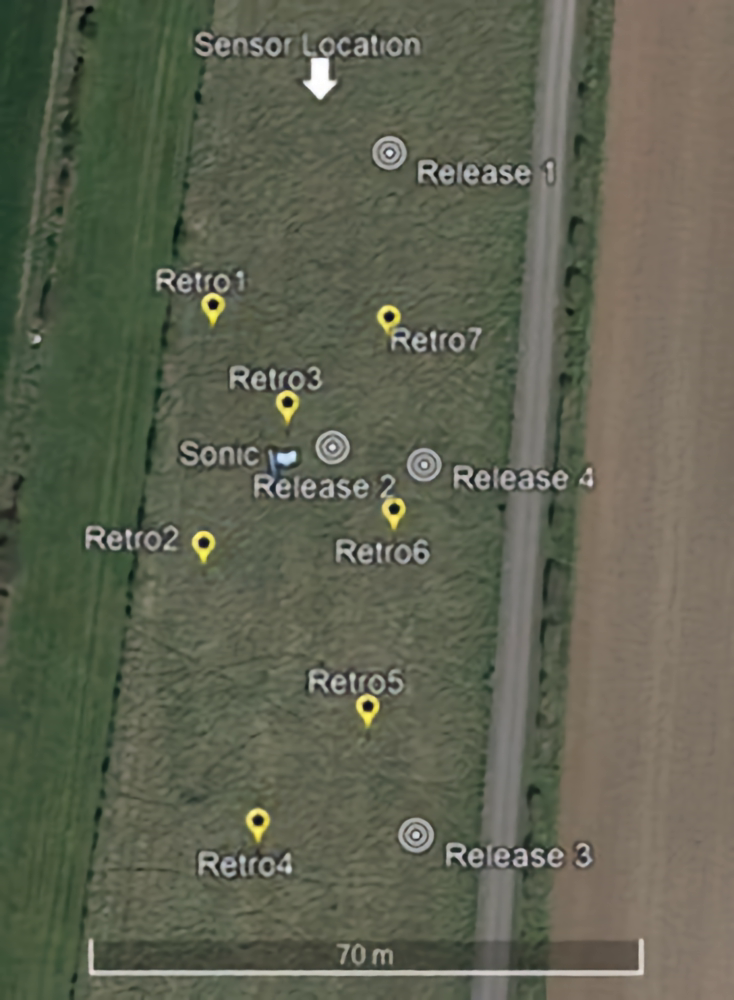}
        \caption{Methane release set up at the Chilbolton observatory, UK. Credit: \cite{hirst2020methane} Supporting Information S1. A laser dispersion spectrometer measures path-averaged CH4 concentrations between the sensor and each retros. ``Sonic" indicates the emplacement of the 3-D ultrasonic anemometer measuring the wind speed and direction. }
        \label{chilboltonplot}
\end{figure}

\noindent
\textbf{\textit{Layout of paper:}} Section \ref{Section2} focuses on the forward model, our framework for simulating atmospheric gas dispersions, unsteady-state wind fields, and ground sensor measurements. We employ the Gaussian plume model for gas dispersion \citep{stockie2011mathematics}, the Ornstein-Uhlenbeck (OU) process for unsteady-state wind fields \citep{uhlenbeck1930theory}, and simulate point sensor measurements to mimic real-world data acquisition scenarios incorporating background concentration. Section \ref{Section3} introduces the core of our inference methodology, we present our parameter estimation method using the Manifold Metropolis adjusted Langevin algorithm (M-MALA) in combination with Gibbs sampling. Section \ref{Section4} presents the results from our simulation case study where we test the robustness of our inversion method for single source cases, under varying atmospheric, data collection, and sensor layout conditions. We also show the importance of estimating wind sigmas when estimating source characteristics. In Section \ref{Section5}, we implement our inversion method on data from a real-world field experimental campaign (see Figure \ref{chilboltonplot}) reported by \cite{hirst2020methane}. This dataset contains wind field and methane measurements for controlled release trials. Finally, Section \ref{Section6} summarizes the paper and suggests potential lines of future work.

\section{Atmospheric Gas Concentration and Sensor Measurements}\label{Section2}
\begin{figure}
    \centering
        \includegraphics[width=0.7\textwidth]{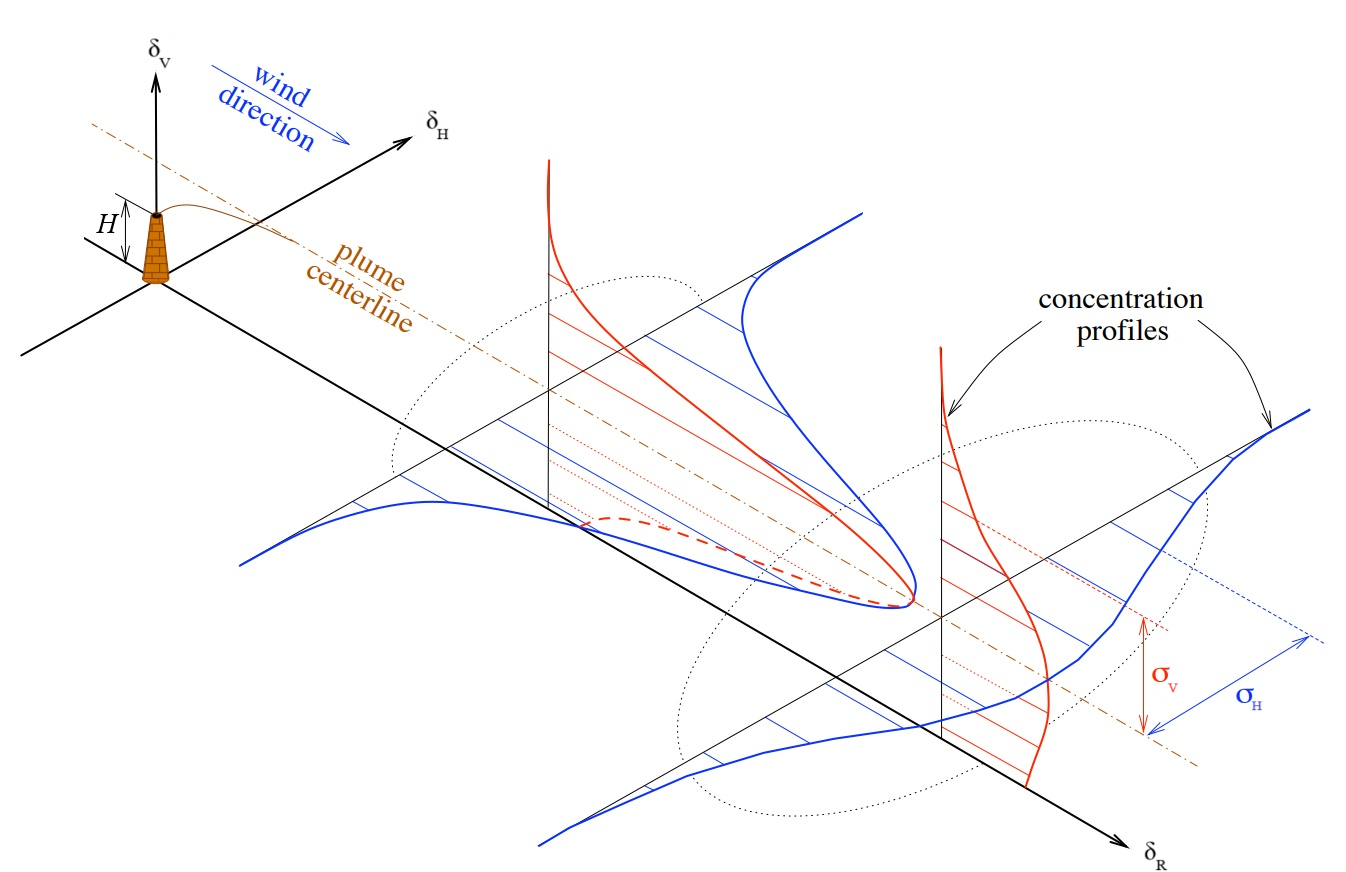}
        \caption{Representation of the Gaussian plume model. Credit: \cite{stockie2011mathematics}.}
        \label{gaussianplumerepresentation}
\end{figure}

In this section, we present the modeling framework of the simulation, incorporating the Gaussian plume model for gas dispersion (Section \ref{Section2.1} \& \ref{Section2.2} ); see Figure \ref{gaussianplumerepresentation} for visual representation, OU process for wind fields (Section \ref{Section2.3}), and sensor measurements when accounting for background gas concentration (Section \ref{Section2.4}). By combining these three elements, we gain a holistic perspective on air quality dynamics, enabling a deeper understanding of pollutant transport.



\vspace{2mm}

\noindent
The formulation of the forward model sets the stage for the subsequent exploration of inversion modeling, where we aim to estimate sources' location and emission rate by leveraging the simulated sensor observations and gas dispersion patterns. We seek to estimate point sources mixed with a spatially varying background concentration.

\subsection{Modeling Gas Dispersions using the Gaussian Plume Model}\label{Section2.1}

A variety of gas dispersion models have been developed, each differing in accuracy and complexity, with three primary categories being prominent. Gaussian plume models, exemplified by ISC3 \citep{osti_422988}, AERMOD \citep{cimorelli2005aermod}, and ADMS 6 \citep{CARRUTHERS1994139}, operate on the assumption of a Gaussian distribution and are widely utilized. Gaussian puff models, such as CALPUFF  \citep{scire2000user}, conceptualize the plume as composed of discrete puffs, while high-fidelity computational fluid dynamics (CFD) models, like Fluidyn-Panache \citep{Libre201137}, employ rigorous numerical techniques.

\vspace{2mm}

\noindent
In practical application, selecting the most suitable model depends on the specific requirements and resources of the modeler. Gaussian plume and puff models are often preferred due to their practicality, especially when comprehensive spatio-temporal wind field data required by CFD models are not readily available to set the initial condition and boundary conditions. Typically, wind data is collected at single points in space, limiting the applicability of CFD models.

\vspace{2mm}

\noindent
The Gaussian plume model is noteworthy for its computational efficiency and straightforward implementation. It is a closed-form analytical expression, that allows simulation of the continuous emission from a single source under the assumption of unidirectional wind flow in an unbounded space. Gaussian plume models have found widespread application in various industries, often serving as a tool for monitoring and regulating emissions from industrial projects. An example of their use can be seen in the work of \cite{LUSHI20101097}, who employed a Gaussian plume model to estimate the emission rates of a large lead-zinc smelting operation in Trail, British Columbia. Similarly, \cite{ramadan2008total} utilized this model to calculate the concentration of sulfur dioxide resulting from existing power stations in Kuwait. These applications demonstrate the practical utility of the Gaussian plume model in assessing and managing the dispersion of pollutants, aiding in environmental impact assessments, urban planning, and emergency responses, among other critical areas.

\vspace{2mm}

\noindent
In this paper, the Gaussian plume model is used to model the dispersion of methane. The Gaussian plume equations are derived from the advection-diffusion equation (\ref{advectiondiffusion}) which is a PDE describing the transport of a substance in three-dimensional space whose mass concentration is represented by a function $C((x,y,z),t)$[kg/m$^3$].

\begin{equation}\label{advectiondiffusion}
    \frac{\partial C}{\partial t} + \nabla  \cdot  (C\boldsymbol{u}) = \nabla  \cdot  (K\nabla C) + S.
\end{equation}

\noindent
$S((x,y,z),t)$[kg/m$^3$s$^{-1}$] provides the source emission rates, $K((x,y,z))$[m$^2$/s] are the diffusion coefficients (from eddy and molecular diffusion), and $\boldsymbol{u}(((x,y,z),t)$[m/s] is the wind velocity field. Using assumptions made by \cite{stockie2011mathematics} and following the derivations in Supplementary Materials A.1, we can write the close-form analytical expression for the Gaussian plume solution as:

\begin{equation}\label{gplume3}
\begin{split}
                c(x,y,z; \tilde{x},\tilde{y},\tilde{z}) &= \frac{10^6}{\rho_{\text{CH4}}}\frac{s}{2 \pi u \sigma_H \sigma_V}\exp{\Bigg\{-\frac{\delta_{H}^2}{2\sigma_H^2}\Bigg\}}  \times  \Biggl(     \exp \left\{ - \frac{\delta_V^2}{2 \sigma_V^2}\right\} \\ &  \qquad + \sum_{j=1}^{n_{\text{refl}}} \Biggl[\exp \Bigg\{ -\frac{1}{2} \frac{(2 \lfloor (j+1)/2 \rfloor P + (-1)^j(\delta_V+H)-H)^2}{\sigma_V^2} \Bigg\}  \\ & \qquad  \qquad +  
    \exp \Bigg\{ -\frac{1}{2} \frac{(2 \lfloor j/2 \rfloor P + (-1)^{j-1}(\delta_V+H)+H)^2}{\sigma_V^2} \Bigg\} \Biggr] \Biggr).
\end{split}
\end{equation}

\vspace{2mm}

\begin{center}
\begin{tabular}{clc}\label{GPtable}
\textbf{Variable} & \textbf{Definition} & \textbf{Units} \\
\hline
$c(x,y,z) \in \mathbb{R}$ & Plume gas concentration at $(x,y,z)$ & PPM \\
$(x,y,z) \in \mathbb{R}^3$ & Three-dimensional coordinates & m \\
$(\tilde{x},\tilde{y},\tilde{z}) \in \mathbb{R}^3$ & Source location  & m \\
$\rho_{\text{CH4}} \in \mathbb{R}^+$ & Methane density & kg/m$^3$ \\
$s \in \mathbb{R}^+$ & Source emission rate & kg/s \\
$u \in \mathbb{R}^+$ & Wind speed & m/s \\
$P \in \mathbb{R}^+$ & Height of the ABL  & m \\
$H \in \mathbb{R}^+$ & Height of the source & m \\
$j \in \mathbb{Z}^{1+}$ & Reflection number \\
$n_{\text{refl}} \in \mathbb{Z}^{1+}$ & Maximum number of reflections \\
$\delta_H$, $\delta_V \in \mathbb{R}$ & Horizontal and vertical offsets & m \\
$\sigma_H$, $\sigma_V \in \mathbb{R}^+$ & Wind sigmas & m  \\

\end{tabular}
\end{center}

\noindent
In the table above we define the atmospheric boundary layer (ABL) as the lowest part of Earth's atmosphere, the behavior of which is influenced by its height and its contact with the Earth's surface (as opposed to the free atmosphere lying above the ABL). We assume the boundary between the ABL and the free atmosphere to be impermeable, reflecting gas emissions within the ABL and keeping them trapped within the lower atmosphere. For the simulation case study (Section \ref{Section4}) and the inversion on the real-world dataset (Section \ref{Section5}) we set $n_{\text{refl}}=3$, based on previous applications experience.

\subsection{Parametrization of the Wind Sigmas}\label{Section2.2}

The Gaussian plume model (\ref{gplume3}) contains atmospheric parameters that influence the shape of the plume, such as the horizontal and vertical wind sigmas, $\sigma_H$, and  $\sigma_V$. These represent the standard deviation of the horizontal and vertical Gaussian distributions for gas concentration which shape the Gaussian plume model. A large literature exists on choosing wind sigmas and originated with Pasquill's ASCs \citep{pasquill}. Pasquill's approach first determines the local ASC using meteorological data, then uses a dispersion scheme to fix wind sigmas according to the ASC. Nowadays, ASC-based dispersion schemes remain popular in practice \citep{kahl2018atmospheric}, with Briggs \citep{briggs1973diffusion}, Smith \citep{pasquill1983atmospheric}, Pasquill-Gifford, and Chinese National Standard being common choices \citep{mao2020impacts}. These power-law dispersion schemes based on downwind distances fix the wind sigma parameters by selecting dispersion parameters from ASC-based tables. However, atmospheric conditions are extremely complex, and by fixing the dispersion parameters we risk misspecifying them \citep{finn2016project}. In this paper, we present a method to estimate the wind sigmas by estimating the dispersion parameters without relying on the ASCs. We generalize the power-law parametrization from \cite{hirst2013locating} by adding dispersion parameters $a_H \in \mathbb{R}^+$, $a_V\in \mathbb{R}^+$, $b_H \in (0,1]$, and $b_V \in (0,1]$. For time $t=1, 2, \cdots, n_T $ and fixed location $(x, y, z)$ :

\vspace{-2mm}

\begin{equation}
    \begin{split}
        \sigma_{H_t} &= a_H \left( \delta_R \tan(\gamma_{H_t})  \right)^{b_H} + w,\\
        \sigma_{V_t} &= a_V \left( \delta_R \tan(\gamma_{V_t})  \right)^{b_V} + h,
    \end{split}
    \label{windsigmadraxler}
\end{equation}

\vspace{2mm}

\noindent
where $\boldsymbol{\gamma_H} \in \mathbb{R}^+$ and $\boldsymbol{\gamma_V} \in \mathbb{R}^+$ are the 1 minute rolling standard deviation of the horizontal and vertical wind direction time series, $\delta_R \in \mathbb{R}^+$ is the downwind distance of location $(x, y, z)$ from the source located at $(\tilde{x},\tilde{y},\tilde{z})$, $w \in \mathbb{R}^+$ is the source's half-width, and $h \in \mathbb{R}^+$ the source aperture's half-height. When the measurement location is upwind from the source we set the Gaussian plume concentration contribution to zero. In Section \ref{Section4}, we show the impact of misspecified wind sigmas on source parameter estimation and how estimating dispersion parameters reduces this bias.

\subsection{Simulating Unsteady-State Wind Field using Ornstein-Uhlenbeck Process}\label{Section2.3}

The OU process is a stochastic process often used to model the behavior of physical systems that tend to revert toward a mean or equilibrium state \citep{uhlenbeck1930theory}. When simulating wind speeds and wind directions, the OU process can be useful for generating realistic, time-varying wind fields. Here we model the wind speed and direction as two separate stochastic processes. Their temporal evolution is modeled using an OU process with mean set to the desired average wind speed and direction. By incorporating the OU process into wind simulation models, it is possible to generate wind fields that exhibit realistic temporal and spatial variability, which is useful for many applications such as wind energy production, air pollution dispersion modeling, and neuronal activity \citep{arenas2020ornstein, boughton1987stochastic, ricciardi1979ornstein}.

\vspace{2mm}

\noindent
The OU process can be numerically simulated using the Euler-Maruyama method \citep{maruyama1955continuous}. The Euler-Maruyama scheme discretizes the OU process into a series of time steps, and the stochastic differential equation governing the process is approximated using a finite-difference equation. We can therefore simulate an OU process numerically with standard deviation $\xi \in \mathbb{R}^+$ and correlation time 
$\Theta \in \mathbb{R}^+$:
\begin{equation}\label{Ornstein-Uhlenbeckequation}
{\displaystyle \eta(t+dt)=\eta(t)-\Theta \,dt\,\eta(t)+  \nu _{t} \xi {\sqrt {2\,dt\,\Theta }}},
\end{equation}
where $\nu _{t}$ is a random number sampled independently at every time-step ${\displaystyle dt} \in \mathbb{R}^+$ from a standard normal distribution \citep{kloeden2002numerical}.

\subsection{Point and Beam Sensor Measurements}\label{Section2.4}

Our inversion model utilizes measurements of atmospheric gas concentration. Different types of gas sensor platforms are available, such as satellites, airplanes, drones, line-of-sight/beam sensors, and point detectors \citep{fox2019review},  with each sensor type having its advantages. Point sensors can provide very high accuracy measurements but have poor spatial coverage whereas satellites can cover vast areas at the cost of measurement precision. This paper focuses on ground sensors (point and beam sensors) since these are the most common techniques for continuous fence line monitoring of assets.

\vspace{2mm}

\noindent
Ground sensors measure gas concentrations over time at fixed locations. Assuming measurement error $\boldsymbol{\epsilon}$ and Gaussian plume model concentrations $\boldsymbol{A}\boldsymbol{s}$, the data collected can be represented by the following equation:
\begin{equation}\label{measurementequation}
    \boldsymbol{d} = \boldsymbol{A}\boldsymbol{s} + \boldsymbol{\beta} + \boldsymbol{\epsilon}.
\end{equation}
The $n_{\text{obs}}$ data points collected are denoted by a $n_{\text{obs}} \times 1$ vector $\boldsymbol{d}$, while $\boldsymbol{A}$ is a coupling matrix with dimensions $n_{\text{obs}} \times n_{\text{src}}$, where $n_{\text{src}}$ represent the number of sources in our model. The elements of $\boldsymbol{A}$ are the Gaussian plume model concentrations at the sensor locations for a unit source emission at the source location. Here, we use a Gaussian plume model, however, more accurate spatial discretization models of the gas dispersion equations are potential alternatives; such as a finite volume method discretization of the advection-diffusion equation (\ref{advectiondiffusion}) \citep{moukalled2016finite, calhoun2000cartesian}.
The vector $\boldsymbol{s}$ has dimensions $n_{\text{src}} \times 1$ and contains the emission rate for each source. The spatially varying and temporally stationary background gas concentration is represented by the $n_{\text{obs}} \times 1$ vector $\boldsymbol{\beta}$, from a Gaussian field with $\boldsymbol{\beta} \sim \text{N}(\boldsymbol{\mu_{\beta}}, \boldsymbol{\Sigma_{\beta}})$, for $\boldsymbol{\mu_{\beta}} \in \mathbb{R}^+$ and covariance matrix $\boldsymbol{\Sigma_{\beta}}$.  Lastly, $\boldsymbol{\epsilon}$ denotes the measurement error vector, where $\epsilon_k \overset{\mathrm{iid}}{\sim}  \text{N}(0, \sigma^2)$ for $k=1,2, \cdots, n_{\text{obs}}$ and $\sigma^2 \in \mathbb{R}^+$.


\section{Probabilistic Inversion for Gas Emission Problems}\label{Section3}

 Building upon Section \ref{Section2}, we now explore the inversion model implemented to estimate the source locations and emission rates. By leveraging the simulated sensor observations and the knowledge of gas dispersion patterns, the inversion model offers a valuable tool for identifying and quantifying the precise source characteristics.

\vspace{2mm}

\noindent
Using $i=1,2, \cdots, n_{\text{src}}$, $j=1,2, \cdots, n_{\text{sns}}$, and $t=1,2, \cdots, n_{T}$ to represent sources, sensors, and observation time points, respectively, for every pair $(j,t)$ we have recorded measurements $\boldsymbol{d}_j = (d_{j1}, d_{j2}, \cdots, d_{jn_T})^T$. Each sensor takes a measurement at time $t$ giving a $n_{\text{obs}} \times 1$ vector of observations $\boldsymbol{d} = (\boldsymbol{d}_1^T, \boldsymbol{d}_2^T, \cdots, \boldsymbol{d}_{n_{\text{sns}}}^T)^T$. Each sensor's concentration measurements are a combination of gas emitted from the sources, background gas concentration, and measurement error variance. The sources' contributions for unit emission rates are denoted by the $n_{\text{obs}} \times n_{\text{src}}$ matrix $\boldsymbol{A}$, modeled using the Gaussian plume equation (\ref{gplume3}). $\boldsymbol{A}_{ki}$ is obtained by computing equation (\ref{gplume3}) for a specified source location $(\tilde{x}_i, \tilde{y}_i, \tilde{z}_i)$ and dispersion parameters $a_H, a_V, b_H, b_V$ from equation (\ref{windsigmadraxler}). Each source has an emission rate denoted $s_i$ used to rescale the coupling matrix $\boldsymbol{A}$. Each sensor's measurements contain a different spatially varying background gas concentration $\boldsymbol{\beta}_j = (\beta_{j1}, \beta_{j2}, \cdots, \beta_{jn_T})^T$ where $ \beta_{j1} = \beta_{j2} = \cdots = \beta_{jn_T}$ giving $\boldsymbol{\beta} = (\boldsymbol{\beta}_{1}^T,\boldsymbol{\beta}_{2}^T, \cdots, \boldsymbol{\beta}_{n_{\text{sns}}}^T)^T$.

\vspace{2mm}

\noindent
We are interested in estimating emission rates $\boldsymbol{s} = (s_1, s_2, \cdots, s_{n_{\text{src}}})^T$ and corresponding source locations $(\boldsymbol{\tilde{x}}, \boldsymbol{\tilde{y}}, \boldsymbol{\tilde{z}}) = ((\tilde{x}_1,\tilde{x}_2, \cdots, \tilde{x}_{\text{src}})^T, (\tilde{y}_1,\tilde{y}_2, \cdots, \tilde{y}_{\text{src}})^T, (\tilde{z}_1,\tilde{z}_2, \cdots, \tilde{z}_{\text{src}})^T )$. These are estimated simultaneously with  $\boldsymbol{\beta}$, $\sigma^2$, and $a_H,a_V,b_H,b_V$  to reduce bias. For simplicity, we fix the sources' height near the ground $\tilde{z}_i \approx 0$.

\vspace{2mm}

\noindent
Inversion modeling is a powerful technique in various scientific disciplines, particularly geophysics and statistics, and aims to infer unknown parameters or variables from observed data. It involves the mathematical formulation of a forward model that simulates the observed data given a set of input parameters. Inversion modeling reverses this process by estimating the most likely parameters that produced the observed data. MCMC methods are frequently employed in inversion modeling for their ability to explore complex, high-dimensional parameter spaces that are otherwise intractable. MCMC is particularly useful when dealing with nonlinear and non-Gaussian problems, providing robust and probabilistic estimates of model parameters while accounting for uncertainties in both data and model assumptions.

\vspace{2mm}

\noindent
Let $\boldsymbol{\lambda} = \{\boldsymbol{s}, \boldsymbol{\tilde{x}}, \boldsymbol{\tilde{y}}, a_H, b_H, a_V, b_V, \sigma^2, \boldsymbol{\beta}\}$, we can write the full posterior distribution of our inversion problem as:

\begin{equation}
    p(\boldsymbol{\lambda} \mid \boldsymbol{d}) \propto p(\boldsymbol{d} \mid \boldsymbol{\lambda})p(\boldsymbol{\lambda}).
\end{equation}

\noindent
The common set of parameters for prior distributions 
used during the simulation case study (Section \ref{Section4}) and the inversion on the Chilbolton dataset (Section \ref{Section5}) are listed in Supplementary Materials B.3.

\subsection{Gibbs Sampling}

Gibbs sampling \citep{geman1984stochastic} is a fundamental technique in MCMC methods particularly advantageous in scenarios where the joint distribution is difficult to sample directly but where conditional distributions are known or can be easily calculated. When the prior and likelihood functions belong to a conjugate pair, the posterior distribution has a known analytical form. This allows for posterior samples drawn by sequentially sampling from the conditional posterior distributions. The parameters $\sigma^2$ and $\boldsymbol{\beta}$ are estimated using Gibbs sampling with the following priors:
\begin{equation}
  \begin{split}
      \sigma^{2} & \sim \text{Inv-Gamma}(a,b), \\
          \boldsymbol{\beta} & \sim \mathcal{N}(\boldsymbol{\mu_{\beta}}, \boldsymbol{\Sigma_{\beta}}),
  \end{split}  
\end{equation}

\noindent 
where $a \in \mathbb{R}^+$, $b \in \mathbb{R}^+$, $\boldsymbol{\mu_{\beta}}$ is set using historical average background gas concentrations and $\boldsymbol{\Sigma_{\beta}}$ is a diagonal matrix. The mathematical derivations of the following conjugate posteriors are provided in Supplementary Materials A.2:

\begin{equation}
    \sigma^{2} \mid \boldsymbol{\lambda} \setminus \{\sigma^{2}\}  \sim \text{Inv-Gamma}\Bigg( \frac{n_{\text{obs}}}{2} + a \ , \ b + \frac{\sum^{n_{\text{obs}}}(\boldsymbol{d} - \boldsymbol{\beta} - A\boldsymbol{s})^{2}}{2} \Bigg),
\end{equation}

\begin{equation}
\boldsymbol{\beta} \mid \boldsymbol{\lambda} \setminus \{\boldsymbol{\beta}\} \sim \text{N} \Biggl( \left( \frac{1}{\sigma^{2}} \mathbb{I} + \boldsymbol{\Sigma_{\beta}}^{-1} \right)^{-1}\left(\frac{1}{\sigma^{2}}(\boldsymbol{d}-\boldsymbol{A}\boldsymbol{s}) + \boldsymbol{\Sigma_{\beta}}^{-1} \boldsymbol{\mu_{\beta}}\right) \ , \left( \frac{1}{\sigma^{2}}\mathbb{I} + \boldsymbol{\Sigma_{\beta}}^{-1} \right)^{-1} \Biggr).
\end{equation}

\subsection{Manifold Metropolis Adjusted Langevin Algorithm Sampling}

Gibbs sampling is only possible when analytical forms of the conditional posterior distribution are available. The emission rates, locations, and dispersion parameters have a nonlinear relationship, making the derivation of the conditional posteriors extremely challenging. In such cases, gradient-based MCMC methods like the Metropolis-Adjusted Langevin Algorithm (MALA) offer a valuable alternative to sample from the posterior \citep{grenander1994representations}.

\vspace{2mm}

\noindent
Considering the structure of the Gaussian plume model, we expect variables to be correlated. Traditional MALA schemes rely on local gradient information, resulting in inefficient sampling in this scenario. Here M-MALA presents compelling advancements over MALA by accounting for these interdependencies using a Riemann metric tensor to adapt to the local curvature of our target distribution \citep{girolami2011riemann, xifara2014langevin}. This ensures a more efficient and accurate sampling procedure which has been shown to work for similar problems \citep{karimi2023high}. Under M-MALA, sampling of $\boldsymbol{\theta}$ is performed by a Metropolis-Hastings (MH) step with the following proposal distribution:

\begin{equation}
    \boldsymbol{\theta}^* \sim \text{N}_n \left( \boldsymbol{\theta}^{(l-1)} + 0.5 \zeta^{(l-1)} G^{-1} \nabla  \log(p(\boldsymbol{\theta}^{(l-1)} \mid \boldsymbol{d})) ,\zeta^{(l-1)}G^{-1}  \right), 
\end{equation}

\noindent
where $l$ is the current MCMC iteration, $n$ is the number of parameters, $ \boldsymbol{\theta}^*$ are the proposed parameter values, $ \boldsymbol{\theta}^{(l-1)}$ are the parameter values in the Markov Chain at iteration $l-1$,  $\zeta^{(l-1)}$ is the step size at iteration $l-1$, and $ G$ is the Hessian matrix.

\vspace{2mm}

\noindent
However, it is essential to note that M-MALA is often computationally expensive due to the calculation of the Hessian scaling as $\mathcal{O}(n^3)$. To address this challenge and enhance computational speed, our code is implemented in JAX \citep{jax2018github}, a library for automatic differentiation and high-performance computing, enabling efficient sampling and gradient computation for large-scale Bayesian inference tasks.

\subsection{Positively Constrained Manifold-MALA-within-Gibbs}

Combining Gibbs sampling with MH algorithms yields a hybrid approach known as MH-within-Gibbs \citep{chib1995understanding}; here we use M-MALA-within-Gibbs. This methodology leverages the strengths of both techniques to efficiently sample from complex posterior distributions, particularly in scenarios with correlated parameters and nonlinear relationships. The pseudo-code for our implementation of M-MALA-within-Gibbs is presented in Supplementary Materials A.3 and the full code is available at the GitHub repository provided at the end of this paper. We employ log transformations to enforce positivity constraints on emission rates and dispersion parameters, ensuring physically realistic parameter values throughout the sampling process. However, we do not enforce $b_H, b_V \leq 1$ in the MCMC scheme; exceptional values observed serve as an indicator of model misspecification.


\section{Simulation Study}\label{Section4}

A simulation study was conducted to assess the performance of our inversion methodology and identify its limitations. The experiments presented in this section help to understand how varying factors impact parameter estimation, demonstrating the robustness of our approach and highlighting the necessary conditions for it to perform optimally. These are fundamental steps towards applying our method to real-world data (see Section \ref{Section5}), where some factors cannot be controlled and the true parameter values are often unknown. In this section, we demonstrate our ability to simultaneously estimate the source emission rate, location, background gas concentration, measurement error variance, and dispersion parameters in single source cases. We then highlight the importance of estimating dispersion parameters by comparing source estimations when dispersion parameters are assumed to be known or estimated. 

\vspace{2mm}

\noindent
In order to simulate the data for all experiments we follow the steps in Section \ref{Section2} and generate realizations of point sensor temporal observations. Parameter estimation was performed using 20,000 M-MALA-within-Gibbs iterations with initialization values set by a coarse grid search on the emission rate and location followed by a Latin hypercube on all parameters. The code is available at the GitHub repository listed below and was run using Python version 3.10.12 on 4 cores Intel® Xeon® Gold 6248R and 16GB RAM. The algorithm uses fixed seed pseudo-random numbers for all MCMC samples to ensure reproducibility of results and comparability between simulations.

\subsection{Single Source Estimation}

\begin{figure}
    \centering
    \begin{minipage}[b]{0.3\textwidth}
        \centering        \includegraphics[width=\textwidth]{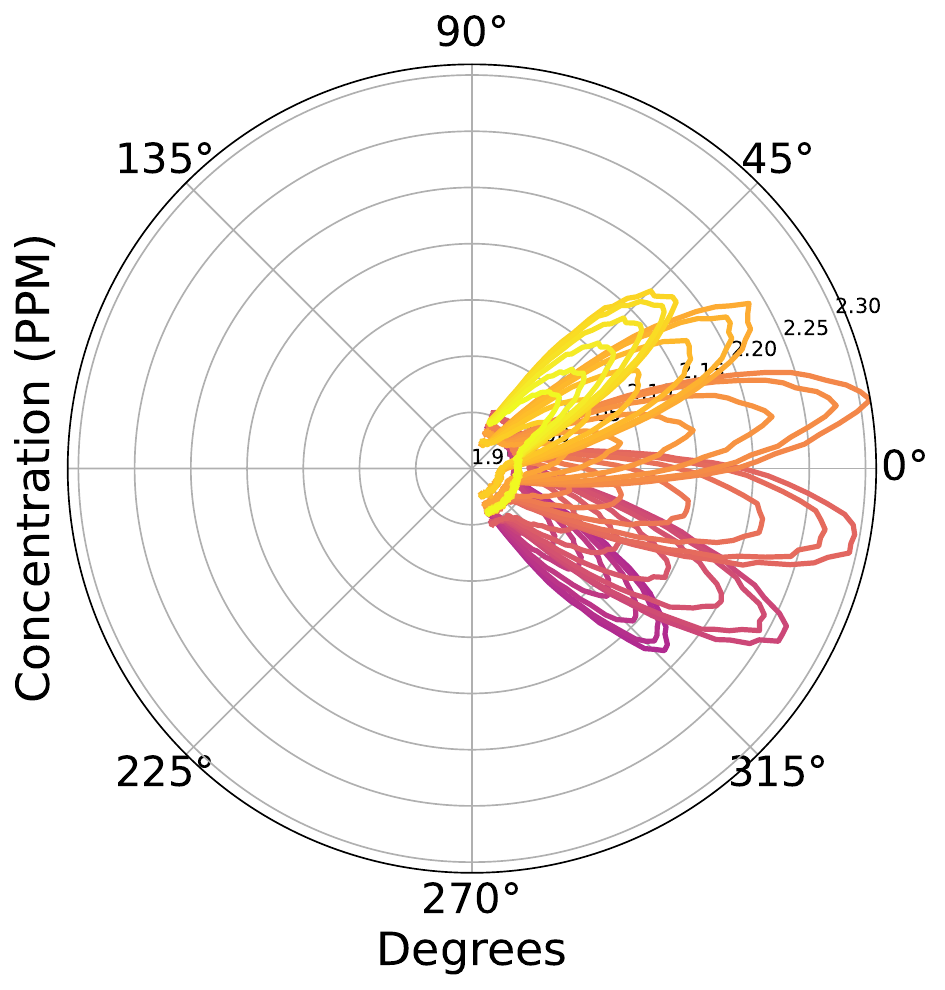}
    \end{minipage}
    \hfill
    \begin{minipage}[b]{0.3\textwidth}
        \centering
        \includegraphics[width=\textwidth]{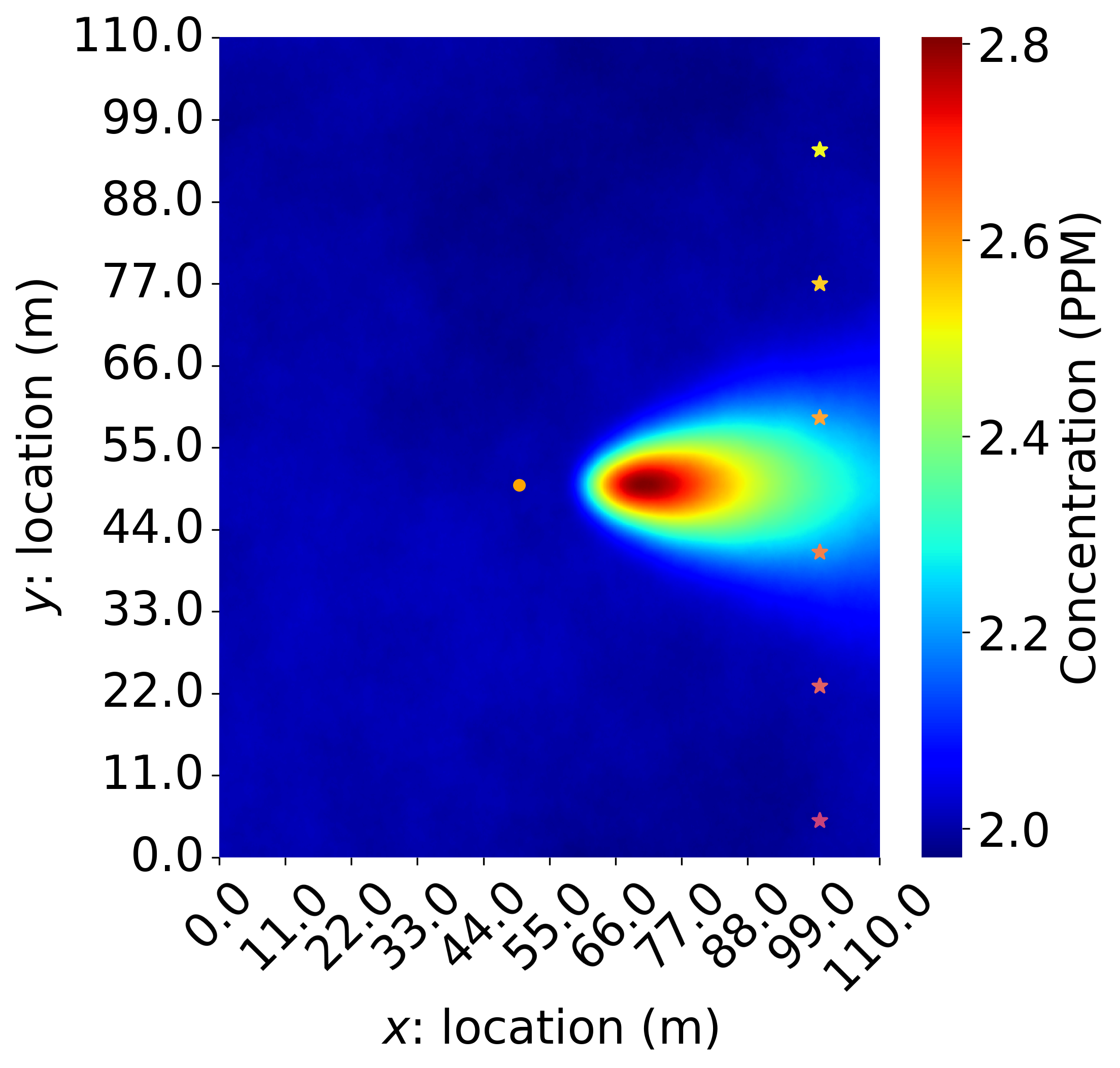}
    \end{minipage}
    \hfill
    \begin{minipage}[b]{0.3\textwidth}
        \centering
        \includegraphics[width=\textwidth]{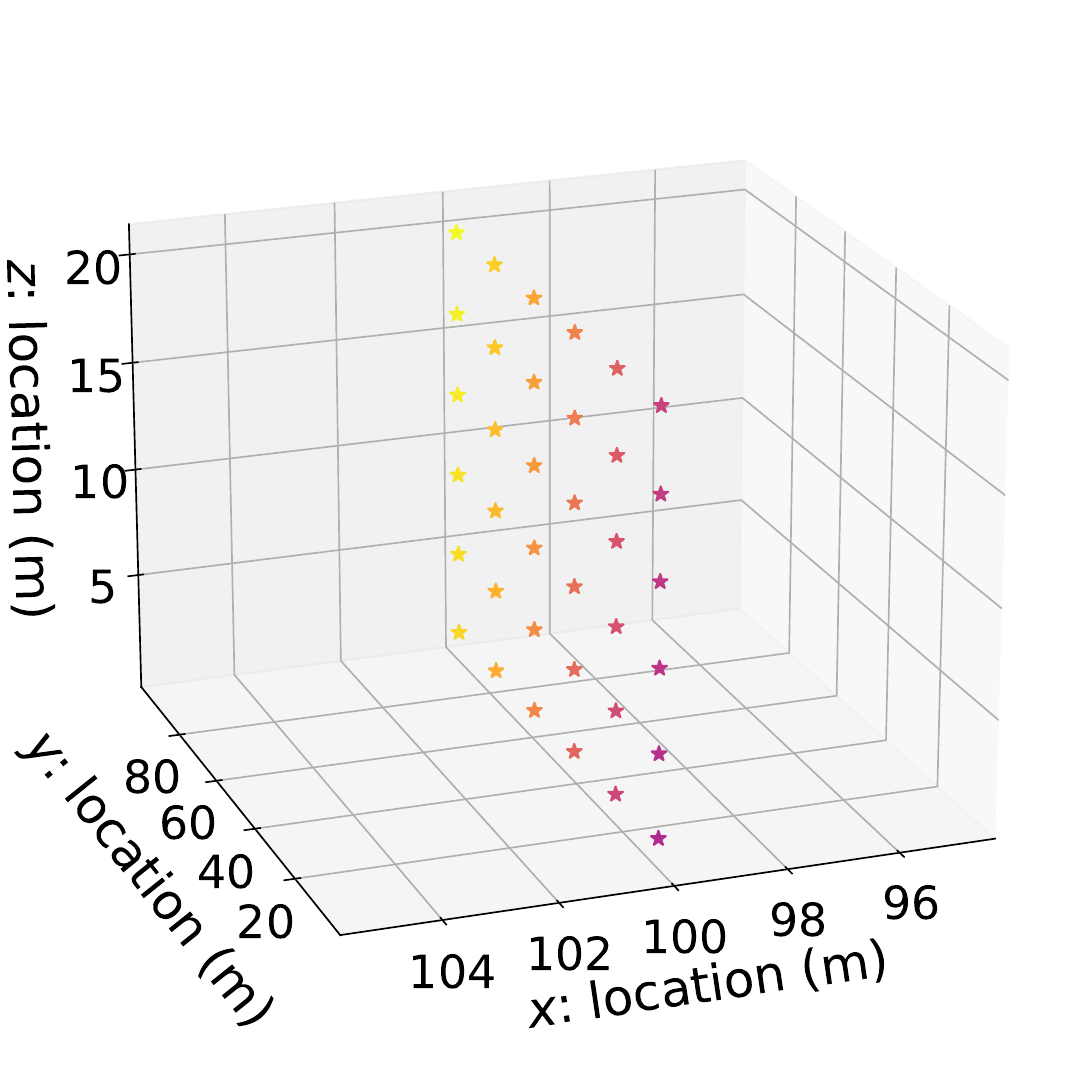}
    \end{minipage}
    \caption{Simulation's level M conditions and sensor layout. The left plot represents the CH4 concentration measurements recorded by the sensors as a function of wind direction. Each line corresponds to a different sensor. The middle plot illustrates the Gaussian plume, background CH4 concentration, and the location of sensors at ground level for wind direction 0°. The right plot represents a grid of 36 evenly spaced point sensors.}
    \label{figure3}
\end{figure}

\begin{figure}
    \centering
        \includegraphics[width=1.0\textwidth]{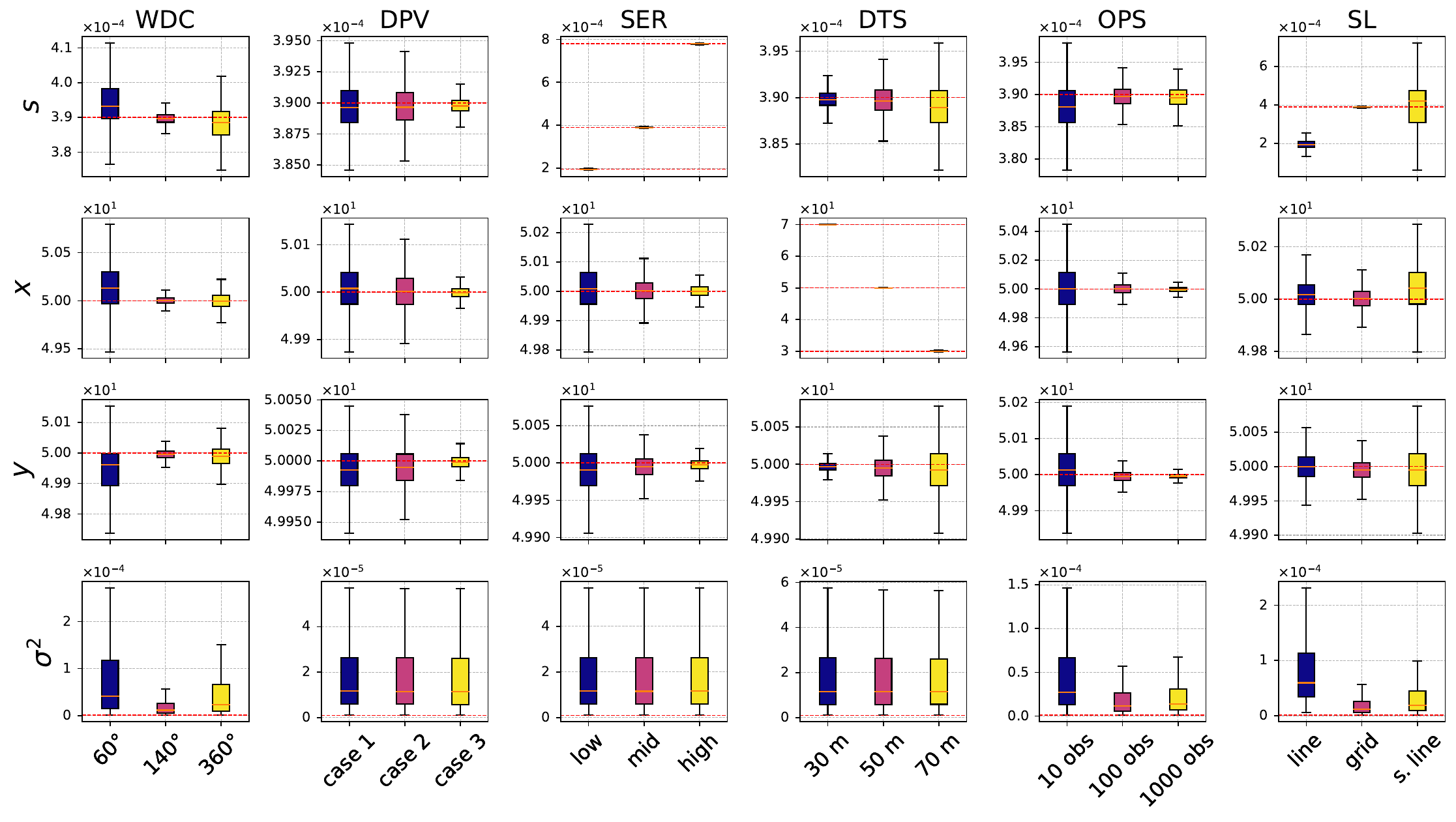}
        \caption{Parameter estimation performance main effects simulation analysis. Each column compares simulations where a single parameter was varied. The column heading indicates the parameter being varied.  WDC: wind direction coverage [degrees°], DPV: dispersion parameter values, SER: source emission rate [kg/s], DTS: distance between the source and sensors [m], OPS: number of observations per sensor, and SL: sensor layout. The rows correspond to the different parameters estimated using M-MALA-within-Gibbs. \textbf{case 1}: $a_H=1.4, b_H=0.9, a_V=1.2, b_V = 0.95$. \textbf{case 2}: $a_H=1.0, b_H=1.0, a_V=1.0, b_V = 1.0$. \textbf{case 3}: $a_H=0.9, b_H=0.8, a_V=0.7, b_V = 0.85$. \textbf{low}: 0.000195 kg/s. \textbf{mid}: 0.00039 kg/s. \textbf{high}: 0.00078 kg/s. 
         \textbf{line}: $36\times1$ line of sensors. \textbf{grid}: $6\times6$ grid of sensors. \textbf{s.line}: $6\times1$ sparse line of sensors. The red dashed lines represent the true values of the estimated parameters.}
        \label{figure4}
\end{figure}

\begin{figure}
    \centering
        \includegraphics[width=1.0\textwidth]{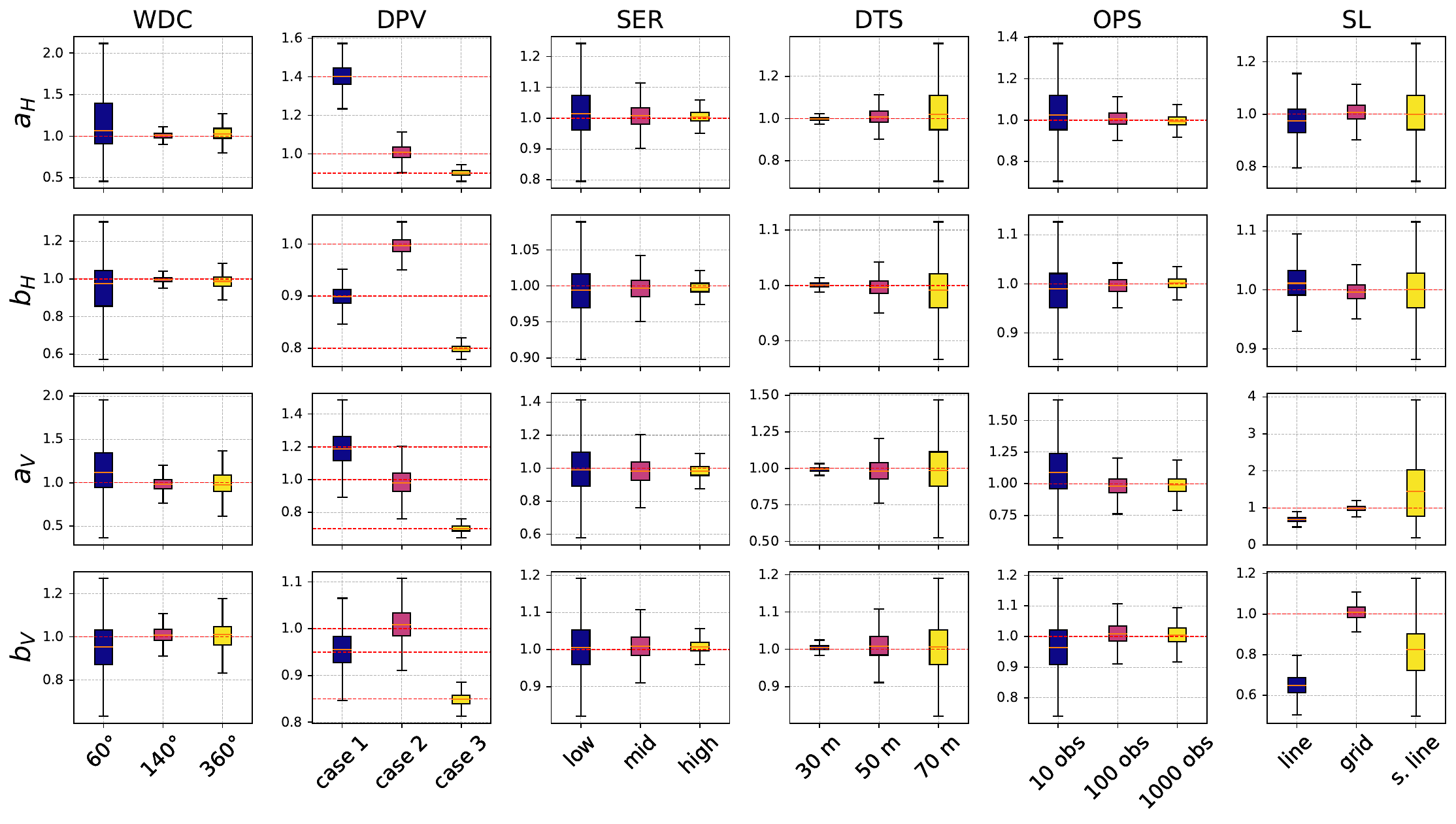}
        \caption{Parameter estimation performance main effects simulation analysis. Each column compares simulations where a single parameter was varied. The column heading indicates the parameter being varied. For the meanings of acronyms in the titles, see caption to Figure \ref{figure4}. The rows correspond to the different parameters estimated using M-MALA-within-Gibbs. \textbf{case 1}: $a_H=1.4, b_H=0.9, a_V=1.2, b_V = 0.95$. \textbf{case 2}: $a_H=1.0, b_H=1.0, a_V=1.0, b_V = 1.0$. \textbf{case 3}: $a_H=0.9, b_H=0.8, a_V=0.7, b_V = 0.85$. \textbf{low}: 0.000195 kg/s. \textbf{mid}: 0.00039 kg/s. \textbf{high}: 0.00078 kg/s. 
         \textbf{line}: $36\times1$ line of sensors. \textbf{grid}: $6\times6$ grid of sensors. \textbf{s.line}: $6\times1$ sparse line of sensors. The red dashed lines represent the true values of the estimated parameters.}
        \label{figure5}
\end{figure}

Simple yet realistic single-source scenarios are useful for examining the inversion capabilities of our model. Our simulations showcase the model's ability to estimate parameters and assess its robustness to parameter variations using the following experimental design. The simulated parameter variations considered are the following: (a) WDC: wind direction coverage in degrees mathematical [°], (b) DPV: dispersion parameter values, (c) SER: source emission rate [kg/s], (d) DTS: distance between the source and sensors [m], (e) OPS: number of observations per sensor,   and (f) SL: sensor layout. These variations assess the robustness of our inversion methodology under different atmospheric conditions (a, b), source characteristics (c, d), and data collection conditions (e, f). For each of the six factors (a)-(f), we define low (L), medium (M), and high (H) levels as detailed in Figure \ref{figure4} and Figure \ref{figure5}. We then perform a ``main effects'' analysis, changing each factor in turn from L to M and then to H, holding all other factors at level M. The level M conditions correspond to an emission source positioned at coordinates (50m, 50m, 5m) within a $110 \text{m} \times 110 \text{m}$ square, emitting at a rate of 0.00039 kg/s (corresponding approximately to the Chilbolton release rates), with all plume dispersion parameters set to 1.0. A grid of 36 evenly spaced sensors positioned downwind of the plume (see Figure \ref{figure3}), collects 100 measurements per sensor at a frequency of 1 Hz and with a measurement error variance of 1e-6 PPM. In practice, sensor layouts will be informed by the local prevailing wind conditions and the physical characteristics of the site. We believe the sensor setup adopted here is useful to explore the role of key design parameters on the quality of inference. An OU process simulates wind speeds with a mean of 6 m/s, and the wind direction varies every second, encompassing a 140° range as depicted in the left plot of Figure \ref{figure3}.  Results of the analysis are shown in Figure \ref{figure4} and Figure \ref{figure5} in terms of box-whisker plots summarizing the marginal posterior distributions of parameters from the MCMC. In all subplots of Figure \ref{figure4} and Figure \ref{figure5}, the middle box-whisker plot
 corresponds to level M for all factors. Detailed results of each inversion presented in Figure \ref{figure4} and Figure \ref{figure5} are available in Supplementary Materials B.1. 

 \vspace{2mm}

\noindent
\textbf{\textit{Varying atmospheric conditions:}} In practice, the wind direction coverage is often positively correlated to the observation period. The longer we collect data the higher the chances of observing a wide range of wind directions. However, a region's prevailing wind can result in narrow wind direction coverage, especially when the observation period is small; e.g. 100 seconds in this simulation. The first column in Figure \ref{figure5} demonstrates the difficulty of estimating dispersion parameters when the wind direction coverage is too small, shown by large uncertainty when the wind direction covers only 60°. However, a full 360° coverage does not lead to optimal inference, due to sensors spending the majority of time outside the plume. The second column contains varying dispersion parameters and shows the model's robustness to different atmospheric conditions. In-depth studies of the impact of varying wind direction coverage are included in Supplementary Materials B.1.1. These reveal the following atmospheric conditions for our inversion method to perform optimally:
\begin{enumerate}
    \item At least one sensor must be in the plume for the majority of the observation period. For a given source location, this is determined by the wind directions and point sensor placements.
    \item The horizontal range of wind directions must exceed the horizontal plume width. This ensures that no point sensor is always in the plume, which makes identification of the dispersion parameters difficult.
\end{enumerate}

\vspace{2mm}

\noindent
\textbf{\textit{Varying source characteristics:}} Source location and emission rate are crucial when monitoring for gas emissions. It is therefore interesting to understand how these affect the inversion capability of our model. From the third column in Figure \ref{figure4} and Figure \ref{figure5}, it is clear that an increase in the emission rate reduces our estimation uncertainty. This is likely due to a more pronounced distinction between the source contribution and the atmospheric background concentration. Similarly, the fourth column indicates a positive correlation between estimation uncertainty and the distance between the source and the sensors.


\vspace{2mm}

\noindent
\textbf{\textit{Varying data collection conditions:}} The fifth column demonstrates the reduction in bias and estimation uncertainty as the sample size increases. The sixth column shows that the sensor layout is a fundamental factor influencing our estimation accuracy. Losing the vertical coverage in the sensor layout has significantly impacted our ability to estimate the vertical dispersion parameters. Due to the structure of the Gaussian plume model, there are positive and negative correlations between the emission rate and the dispersion parameters. To explain the observed gas measurements, a trade-off exists between the dispersion parameters and the emission rate. The former can narrow/widen the shape of the plume while the latter is decreased/increased to explain the measured concentrations. This identifiability issue is shown to be significantly influenced by the sensor layout, we explore it in more detail in Supplementary Materials B.1.6. In the next section we illustrate the bias in source estimation when dispersion parameters are misspecified, highlighting the importance of correctly estimating them.

\vspace{2mm}

\noindent
Overall, the model and inversion methodology presented in Sections \ref{Section2} and \ref{Section3} have demonstrated the ability to estimate all parameters simultaneously and have shown general robustness to the changing atmospheric, source, and data collection conditions applied. However, both wind direction coverage and sensor layout indicate potential limitations of our approach in practice. Dispersion parameters become difficult to estimate when wind direction coverage is small or in the absence of a vertical sensor layout.

\subsection{Estimating Dispersion Parameters}\label{Section4.2}

\begin{figure*}
    \centering
        \includegraphics[width=1\textwidth]{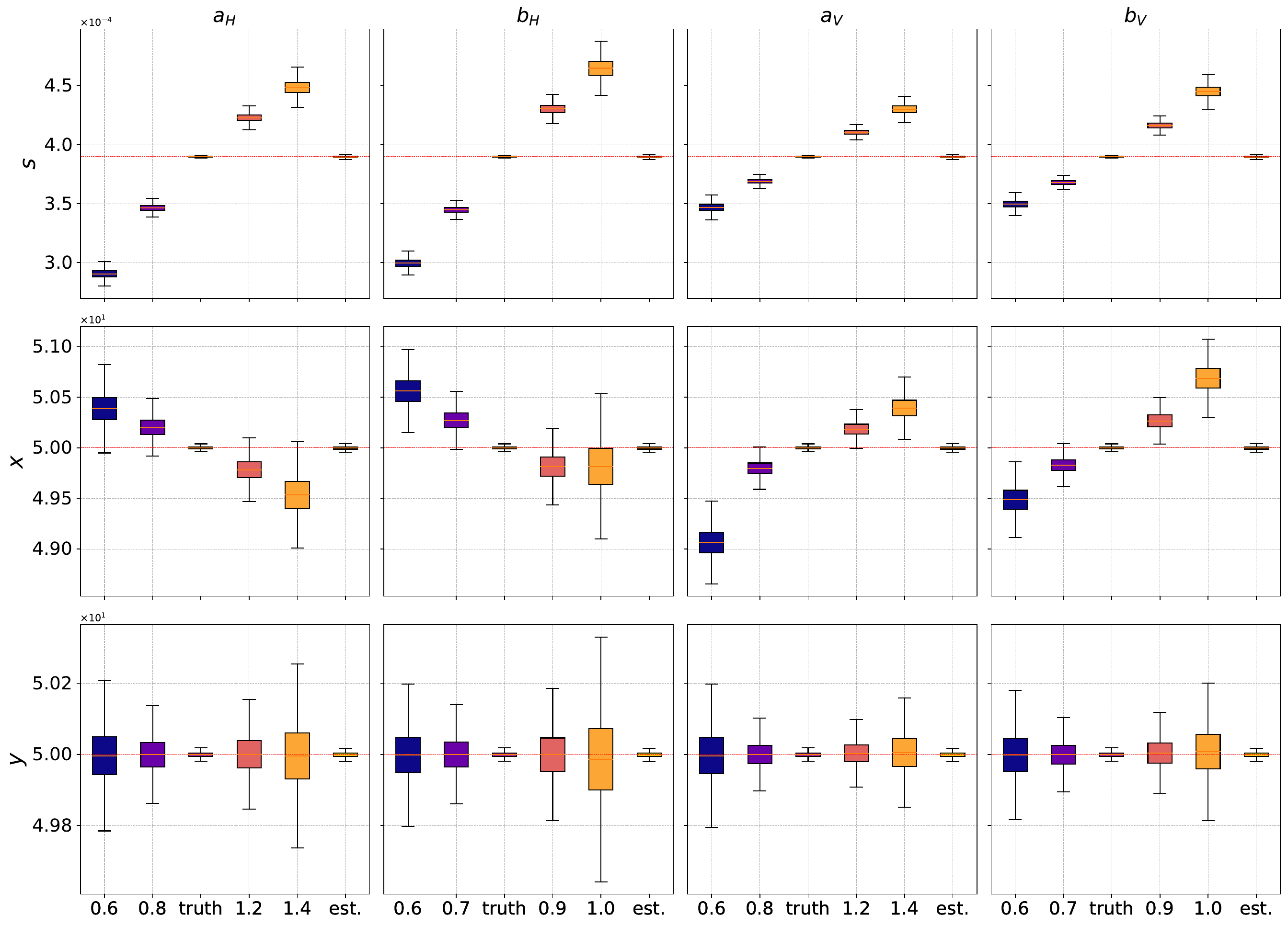}
        \caption{Quantification of dispersion parameter misspecification impact on source estimations. The true values of the estimated parameters are represented by the red dashed lines. The columns correspond to the dispersion parameters being misspecified, with the x-axis indicating the value chosen when misspecifying. These are compared to source estimations when the true dispersion parameter values are chosen, ``truth'', and when all dispersion parameters are being estimated ``est.''. }
        \label{figure6}
\end{figure*}

We now focus on a significant limitation of many gas inversion methods when applying them to real data, the misspecification of dispersion parameters. Similar work was done by \cite{cartwright2019bayesian} who tuned the wind sigmas by estimating a horizontal and a vertical scaling parameter. The parametrization of the wind sigmas used in their work also uses four dispersion parameters, however, these are taken from ASC-based tables, and the ASC was determined using a Monin–Obukh length and an effective roughness length. This approach fixes the dispersion parameters and assumes that rescaling the pairs $\{a_H, b_H\}$ and $\{a_V, b_V\}$ can sufficiently improve the Gaussian plume model. We propose additional flexibility by allowing all four dispersion parameters to be directly, individually, and jointly estimated. Therefore removing bias introduced by the ASC, Monin-Obukh length, and effective roughness length. 
To the best of our knowledge, there is currently no method other than the one we propose in this paper that simultaneously estimates source location, emission rate, background concentration, measurement error variance, and dispersion parameters. In practice, it is common for dispersion parameters to be chosen based on the local ASC. However, as shown in Figure \ref{figure6}, dispersion parameter misspecification can introduce substantial bias in the source estimation. We estimated the source location, emission rate, background concentration, and measurement error variance for various dispersion parameter misspecifications to quantify this bias. We misspecified each dispersion parameter at a time, $a_H, a_V, b_H, b_V$, while fixing the remaining three to their true value. This enables us to identify the main effect biases of each dispersion parameter. Moreover, this experimental design reflects an optimistic reality where three of the four dispersion parameters are correctly specified. In practice, we expect all four ASC-based dispersion parameters to be misspecified because the tables these come from discretize the dispersion parameters when these are in fact continuous. Additionally, the meteorological data required to correctly identify the local ASC is not always available. Parameter estimation based on misspecified dispersion parameters is compared to ``est.": where all four dispersion parameters are estimated simultaneously, and to  ``truth": the unrealistic scenario where all four dispersion parameters have been fixed to the truth. The true values of the dispersion parameters in this simulation are $a_H=1.0, \ a_V=1.0, \ b_H=0.8, \ b_V=0.8$, and all other conditions are set to level M.


\vspace{2mm}

\noindent
From Figure \ref{figure6}, it can be observed that when $a_H$ is too small, or too large, the emission rate is under or overestimated, the same is true for $a_V$. The estimated source distance to the grid of sensors is overestimated when the horizontal dispersion parameters are too small or the vertical dispersion parameters are too large. Similarly, the distance is underestimated when horizontal dispersion parameters are too large and vertical dispersion parameters are too small. However, estimation of the source location coordinate $y$ is robust to misspecification due to the sensors and source layout  (see Figure \ref{figure3}). There is no bias in its estimation but a reduction in uncertainty when dispersion parameters are correctly specified. 
\vspace{2mm}

\noindent
Overall, the misspecification of dispersion parameters shows a strong bias in estimating source characteristics. Meanwhile, estimating the dispersion parameters significantly reduces this bias, as shown by Figure \ref{figure6} where the ``est"  and ``truth" box-whisker plots are almost identical.

\section{Chilbolton Real Data}\label{Section5}

\begin{figure}
        \centering
        \includegraphics[width=0.6\linewidth]{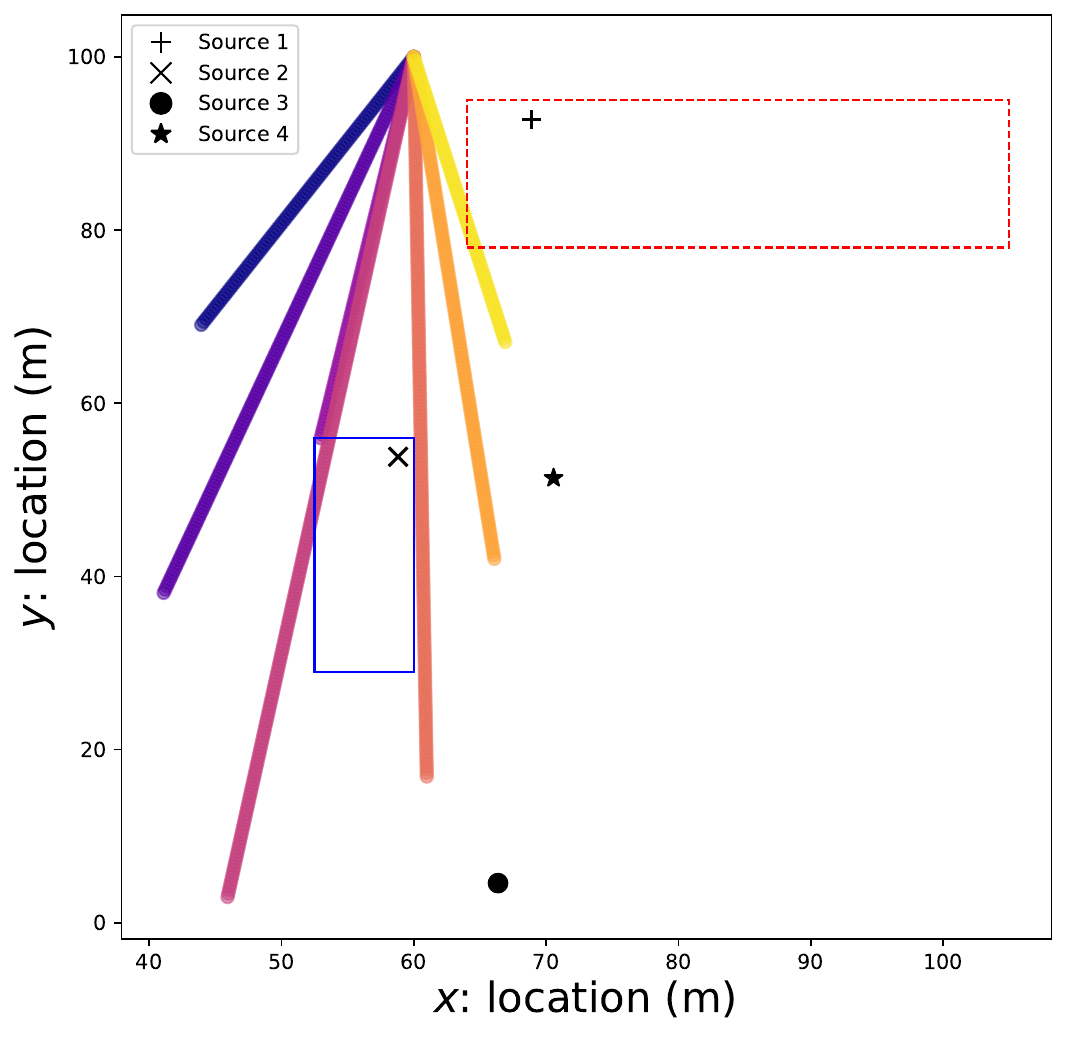}
        \caption{Sensor, beam, and source positions for the Chilbolton experiment. Each colored line corresponds to a different sensor-reflector path. The straight blue line and dashed red line boxes correspond respectively to the plotting area of Source 1 and Source 2 in Figure \ref{figure9}.}
        \label{figure7}
\end{figure}

We now apply our inversion method to the Chilbolton dataset for controlled methane releases made on the flat terrains of the Chilbolton Observatory, Hampshire, UK. As these are controlled releases, the true source locations, and emission rates are known. This enables comparability of our results against other methods applied on the same dataset \citep{hirst2020methane, voss2024multi}. The data for these controlled releases were collected using a multiple open-path laser dispersion spectrometer (LDS) and a single 3D ultrasonic 20Hz anemometer. The LDS measures path-averaged methane concentrations along beams between the LDS and seven fixed reflectors (see Figure \ref{figure7}), with continuous sequential scanning of all beams taking 3s to cover all reflectors. To ensure compatibility between the coupling matrix's point location concentration structure and the beam's path-averaged measurements, 40cm spaced point locations were created along the beams, across which the coupling matrix concentrations were averaged. The Chilbolton experiments contain three controlled releases, two with single sources (Source 1 and Source 2), and one with two sources (Source 3 and Source 4) which we tackle using an extended two source methodology. Each release event includes multiple sub-releases to increase wind direction coverage. Sources were created by 2m$\times$2m aluminum frames laid on the ground, evenly perforated with 1cm spaced holes.  The Gaussian plume model is a representation of the long-term time-averaged concentration under steady-state wind conditions. Thus, over short time scales, it can be a poor representation of the observed data. Averaging the data over longer time periods can improve the correspondence between the model and the (averaged) data. Consequently, measurements from each beam, taken every 3 seconds, were aggregated within 1-minute intervals before estimating the parameters. See Supplementary Materials B.2.1 for details regarding data processing. 

\vspace{2mm}

\noindent
We used four wind sigma parametric forms to implement the inversion procedure presented in Section \ref{Section3} and compared their estimation accuracy. 1) ASC-based dispersion parameters using the Briggs scheme, 2) ASC-based dispersion parameters using the Smith scheme, 3) estimating dispersion parameters using the Smith scheme, and 4) estimating dispersion parameters using Equation (\ref{windsigmadraxler}). From now on, we refer to 3) and 4) as ``estimated Smith" and ``estimated Draxler" respectively. The Chilbolton dataset does not contain the necessary meteorological data to identify the local ASC reliably. The only information available is a picture of the Chilbolton site taken during the releases and the recorded wind speeds.  Based on Table 3 in Supplementary Materials 2.4, the wind speeds recorded and the grey sky in the picture suggest an ASC B or C. However, this approach is not precise enough to confidently determine a single local ASC as required for ASC-based wind sigma parametrizations. Hence, we performed an exploratory data analysis to overcome the lack of meteorological evidence needed to select a unique ASC. This involved comparing the Smith and Briggs-based model predictions of spatial gas concentration measurements to the real data. Details are included in the Supplementary Materials 2.4. The results showed significant differences between the ASC-based model predictions, however, no model stood out as best approximating the observed data. Therefore, these results cannot justify the preference for a specific model. Furthermore, the local ASC could not be determined as the equally most accurate predictions came from the Smith B, Smith C, Briggs A, and Briggs B schemes. This means that we cannot select a unique ASC for our inversion. We therefore perform the inversion using the Smith and Briggs schemes for all available ASCs. The inability to confidently choose an ASC and the bias introduced when incorrectly selecting the ASC underlines the advantages of estimating the dispersion parameter over ASC-based methods. See Supplementary Materials B.2.2-B.2.4 for further details regarding the Smith and Briggs schemes, and ASC determination.

\vspace{2mm}

\noindent
For each model, we ran 10,000 and 5,000 iterations of the M-MALA-within-Gibbs for Sources 1 and 2 respectively,  with a 4,000 and 1,000 burn-in which took on average 27 and 30 minutes for ASC-based models and 48 and 54 minutes for estimated Smith and  Draxler approaches. MCMC convergence was evaluated by investigating traceplots of estimated parameters. Convergence diagnostics are shown in Supplementary Materials B.2.5 and B.2.6. Based on equivalent diagnostics, we found that the MCMC of the multiple source scenario (Source 3 and Source 4) failed to converge after 50,000 iterations taking 1631 minutes. For that reason, we excluded this scenario from the results. A simulation of Chilbolton's Source 3 and Source 4 release, included in Supplementary Materials B.2.7, indicates that the lack of convergence is likely due to insufficient wind direction coverage in the real data (See Supplementary Materials B.2.7.1) Figure \ref{figure8} compares the models' estimation of Source 1 and Source 2 emission rate and location. As expected from the exploratory data analysis, we observe a significant difference between the estimation of ASC-based models. For both Sources 1 and  Source 2, estimation accuracy decreases as ASC moves away from ASC A. Furthermore, Briggs A seems to outperform all other models including the estimated Smith and estimated Draxler models. However, both estimated Smith's and estimated Draxler's estimations are close to the truth, comparable to ASC B-based models, and outperform ASC C, D, E, and F-based models. Figure \ref{figure9} helps to visualize the location estimation in relation to the true source locations and beams. Finally, we examine a model selection criterion and a performance indicator across all estimated variables. Tables \ref{BIC} and \ref{RMSE} containing Bayesian information criterion (BIC) \citep{schwarz1978estimating} and root mean square error (RMSE) quantify each model's goodness of fit and residuals. Only the best ASC-based Smith and Briggs models were presented for clarity. Supplementary Materials B.2.5 and B.2.6 contain details regarding all other ASC-based models' inversion results. According to these tables, Source 1 is best estimated using estimated Smith but Smith B is preferred when penalizing for model complexity, and for Source 2 estimated Draxler is the best fitting model and produced the lowest RMSE.

\vspace{2mm}

\noindent
We can draw two major conclusions from these findings. ASC-based Gaussian plume models can accurately estimate the source characterization on flat open terrain when the local ASC is correctly identified. However, correct identification of the local ASC is not always possible and multiple choices of ASCs may often seem reasonable. The misspecification of the local ASC leads to bias in our estimation which can result in poor estimation accuracy. Therefore, ASC-based models are vulnerable to inaccurate determination of the local ASC, and we believe most practical applications are susceptible to making this mistake. Estimating dispersion parameters is a solution to this problem, it eliminates the bias introduced by specifying an ASC and has shown to be robust and able to accurately estimate source characteristics.

\begin{table}[ht]
\centering
\caption{Comparing models' inversion performances for Source 1 using BIC and RMSE. Optimal models for each performance measure are given in bold. ``Est. Smith" and ``Est. Draxler" estimate the dispersion parameters instead of using ASC-based tables.}
\begin{tabular}{lcccc}
    \toprule
    & \textbf{Briggs A} & \textbf{Smith B} & \textbf{Est. Smith} & \textbf{Est. Draxler} \\
    \midrule
    \textbf{BIC} & 2048 & \textbf{2014} & 2025 & 2058 \\
    \textbf{RMSE} & 0.629 & 0.617 & \textbf{0.611} & 0.622 \\
    \bottomrule
\end{tabular}
\label{BIC}
\end{table}

\begin{table}[ht]
\centering
\caption{Comparing models' inversion performances for Source 2 using BIC and RMSE. Optimal models for each performance measure are given in bold. ``Est. Smith" and ``Est. Draxler" estimate the dispersion parameters instead of using ASC-based tables.}
\begin{tabular}{lcccc}
    \toprule
    & \textbf{Briggs A} & \textbf{Smith C} & \textbf{Est. Smith} & \textbf{Est. Draxler} \\
    \midrule
    \textbf{BIC} & 4386 & 4343 & 4223 & \textbf{3539} \\
    \textbf{RMSE} & 0.577 & 0.611 & 0.552 & \textbf{0.477} \\
    \bottomrule
\end{tabular}
\label{RMSE}
\end{table}

\begin{figure}
        \centering
        \includegraphics[width=\linewidth]{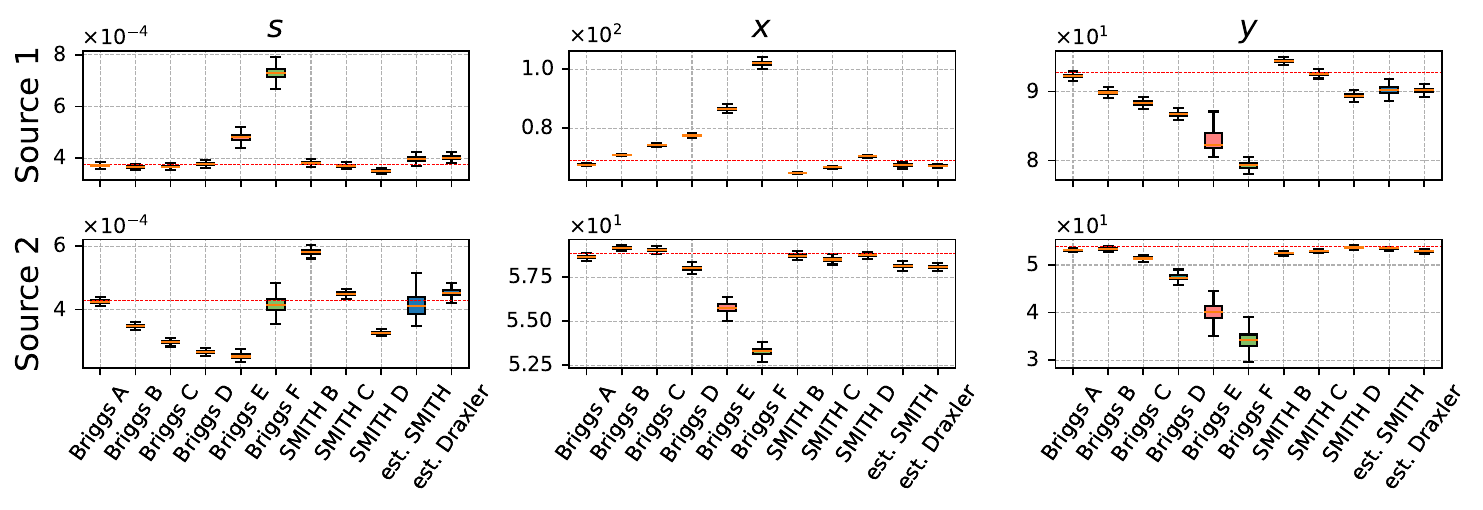}
        \caption{Source 1 and Source 2 emission rate and location estimations for all ASC-based models tested, estimated Smith, and estimated Draxler. The red dashed line represents the true sources' location and rate. }
        \label{figure8}
\end{figure}

\begin{figure}
    \centering
    \begin{minipage}[b]{0.45\textwidth}
        \centering
        \includegraphics[width=\textwidth]{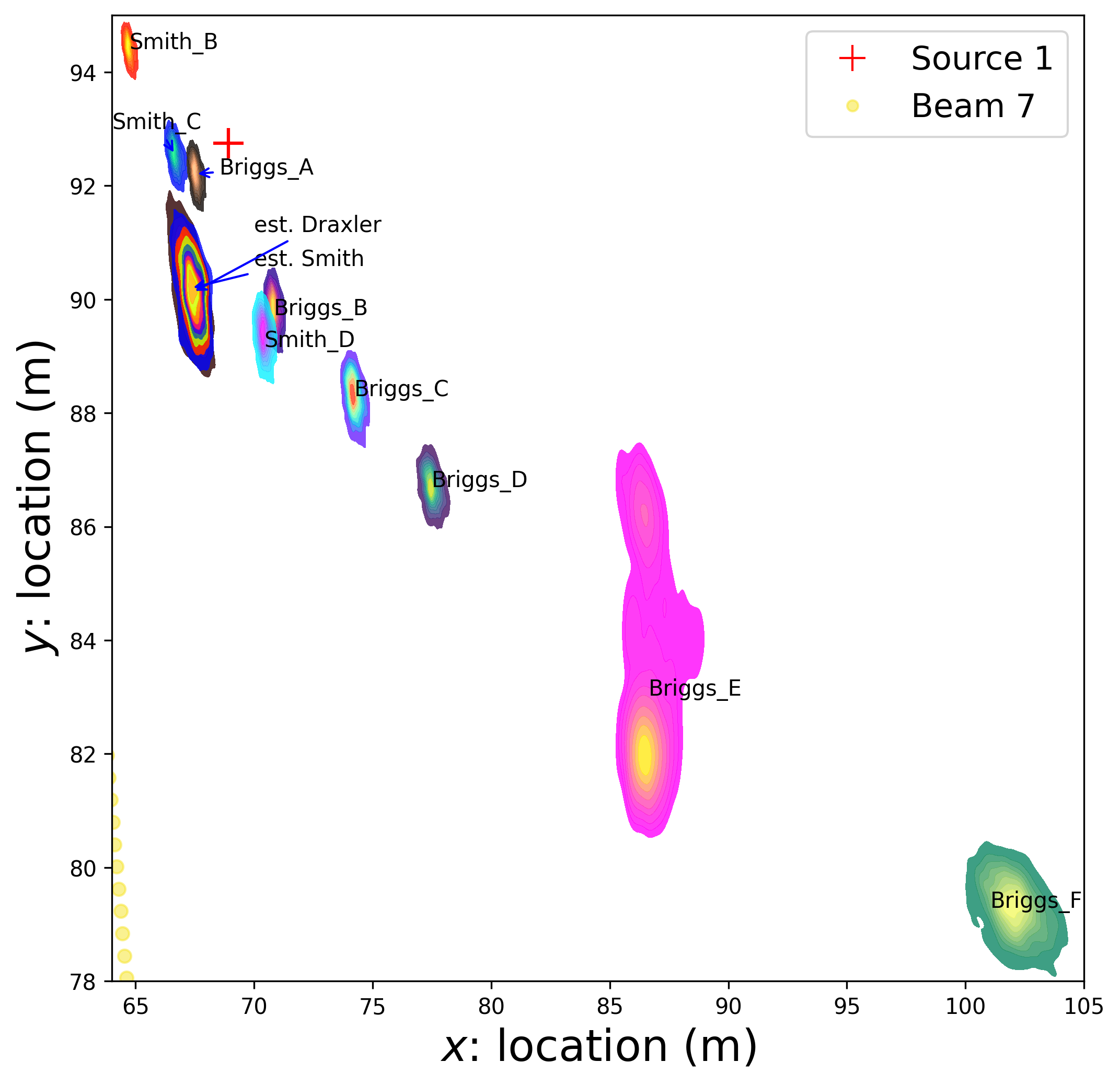}
    \end{minipage}
    \begin{minipage}[b]{0.45\textwidth}
        \centering
        \includegraphics[width=\textwidth]{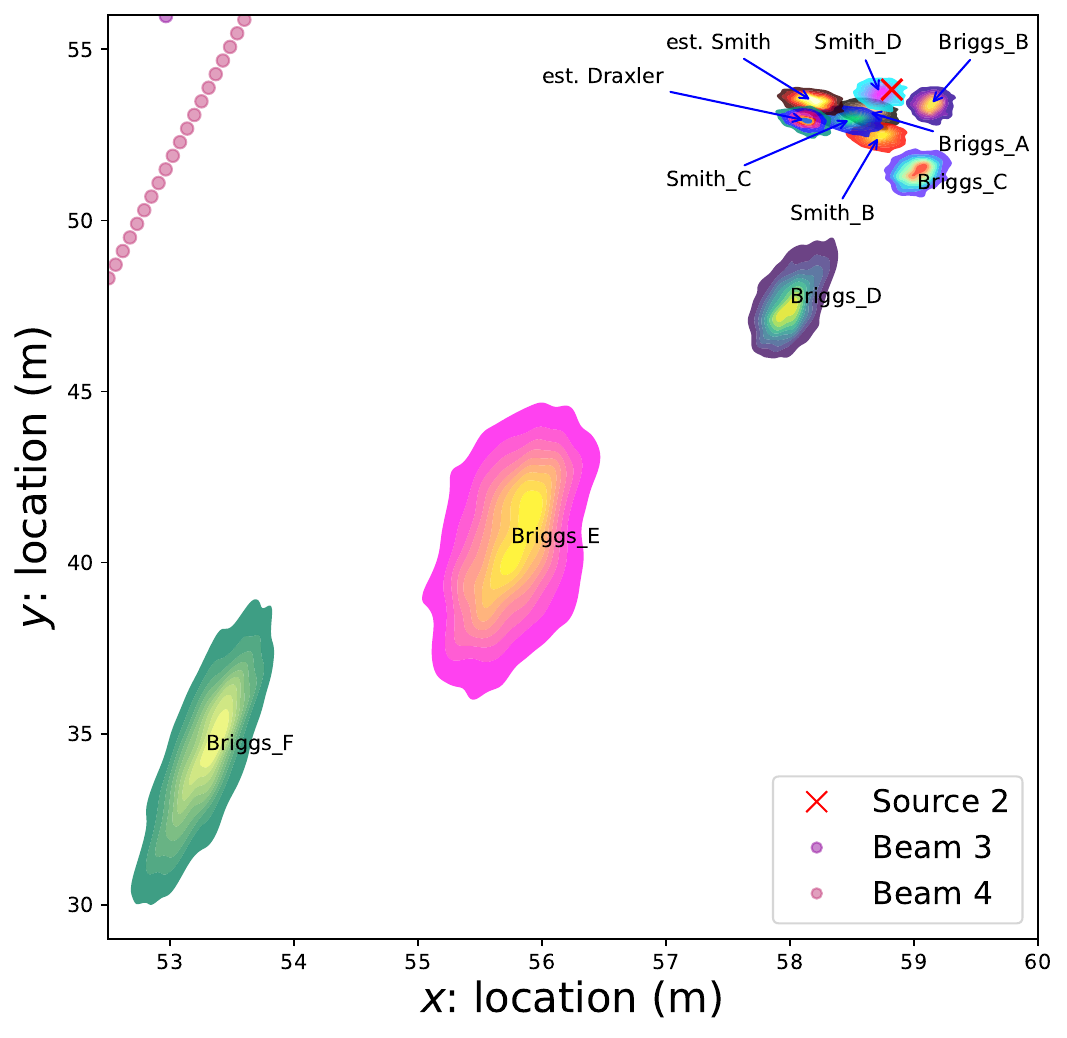}
    \end{minipage}
    \caption{Source 1 (left) and Source 2 (right) location estimation density contours from all ASC-based models tested, estimated Smith, and estimated Draxler. Note: different scales for x and y axis.}
    \label{figure9}
\end{figure}

\section{Discussions}\label{Section6}

We consider a Gaussian plume-based forward model for atmospheric gas dispersion simulating realistic source emissions monitoring scenarios. We also propose an MCMC-based inversion method to estimate the source emission rate, location, gas background concentration, measurement error variance, and dispersion parameters. Our results show that estimating the dispersion parameters reduces bias relative to inference using ASC-based models. When meteorological data is insufficient to determine the stability class,
our model provides robust and accurate results. The dispersion parameters of the Gaussian plume model should therefore be estimated rather than fixed when possible,  ideally incorporating meteorological and methane concentration measurement information. The simulation study in Section \ref{Section4} shows that this is only feasible when data are recorded using an appropriate sensor layout and sufficient wind direction coverage. A large literature exists demonstrating the importance of appropriate sensor layouts in inverse problems \citep{liu2022optimal, liu2022sensor, dia2024greedy}. These conditions ensure optimal inversion estimation and should therefore serve as guidelines when installing ground sensors and collecting data for monitoring purposes. Additionally, the simulation study demonstrated the robustness of our inversion methodology to a wide range of atmospheric, source, and data collection conditions. Finally, in Section \ref{Section5} we proved the effectiveness of our method in practice by applying it to real data.

\vspace{2mm}

\noindent
This paper serves to demonstrate the importance of carefully choosing the dispersion parameters when performing inference in practice. As such, it does not focus on creating the most realistic CFD model, and we could consider the following simple modifications. The observation equation (\ref{measurementequation}) can be extended to model the background gas concentration, the sensor measurement errors, and the gas dispersion in a more physically realistic way. For example, background gas concentration $\boldsymbol{\beta}$ could be modeled using a spatio-temporal Gaussian process or Gaussian Markov random field, potentially accounting for wind field-induced dependence. Under this assumption, it is important to jointly estimate $\boldsymbol{\beta}$ and the measurement error variance $\sigma^2$. Additionally, the assumption that sensor measurement errors are independently and identically normally distributed may be overly simplistic. We might relax this assumption by choosing to model serially correlated errors and modeling sporadic error spikes. Finally, the coupling matrix $\boldsymbol{A}$ could be computed using a more physically realistic forward model able to account for obstacles in the flow field. Traditional finite element or finite volume methods have been extensively studied for this purpose and numerous high-quality software libraries are available e.g. OpenFOAM \citep{OpenFOAM} and ANSYS Fluent \citep{AnsysFluent}. However, solving the PDE every time the estimated parameters change can be time-consuming. Recent advancements in forward model emulators might offer a solution. Physics-informed neural networks (PINNs) have shown great accuracy at solving general nonlinear PDEs \citep{raissi2019physics, cai2021physics} including the advection-diffusion equation \citep{pang2019fpinns, salman2022deep} and 2D Navier-Stokes equations \citep{brahmachary2024unsteady}. Training a PINN can be very computationally expensive and time-consuming. However this cost is amortized, once the training is complete, the solution evaluation is fast \citep{cuomo2022scientific}. This efficiency is particularly advantageous in inversion scenarios where the same PDE must be solved repeatedly with different parameter values.

\vspace{2mm}

\noindent
In the current work, the source is assumed to be located near the ground $\tilde{z} \approx 0$, appropriate for many applications. This assumption can be relaxed by estimating $\tilde{z}$, allowing our methodology to estimate off-ground source characteristics. The source horizontal and vertical half widths $\{h, w\}$ can be estimated similarly; additionally, assuming a spherical source remains a sufficient approximation in many scenarios.

\vspace{2mm}

\noindent
Numerous extensions of the current work are possible. A grid-based version of our method was originally considered, with the center of the grid cell serving as a potential source location. In practice, the number of sources is expected to be small compared to the number of cells; our methodology therefore incorporated a spike and slab prior on the emission rates to constrain the number of cells corresponding to sources. However, this method was abandoned due to its computational cost. Assuming cell-centered sources introduces bias in the parameter estimation, which can be reduced by increasing the grid's resolution. Unfortunately, using a fine grid creates a high-dimensional inversion problem with strong correlations between parameters. Nonetheless, we believe it would be interesting to use the grided approach, with a computationally cheaper non-Hessian-based MCMC method, to identify the number of sources and their emission rates. These estimates could then be used as starting solutions for our inversion method. See for example \cite{pyelq} and \cite{hirst2013locating}.

\vspace{2mm}

\noindent
Finally, for practical online source monitoring applications, it is critical to accommodate temporal variations in source, background, and dispersion parameter characteristics, due to effects of e.g. weather conditions, human activities, and seasonality. To account for these temporal variations, we believe extending our work to state-space models whilst enforcing source sparsity would be an exciting area of research. \cite{voss2024multi} have presented promising results for such approaches on the Chilbolton dataset.


\section*{Data Availability Statement}
The raw Chilbolton data that support the findings of this study are
openly available at the following: \url{https://edata.stfc.ac.uk/items/5c88d121-0e19-4840-a26b-499dba49419a}.


\section*{Code Availability and Supplementary Materials}
Code and data for replicating the study results are available at \url{https://github.com/NewmanTHP/Probabilistic-Inversion-Modeling-of-Gas-Emissions}. The Python package \textbf{sourceinversion} implementing the proposed method and Supplementary Materials A and Supplementary Materials B for this paper are also available on the same GitHub repository.


\section*{Acknowledgements}
   The work was completed while Thomas Newman was part of the EPSRC-funded STOR-i center for doctoral training (grant no. EP/S022252/1), with part funding from Shell Information Technology International Ltd. Christopher Nemeth gratefully acknowledges the support of EPSRC grants EP/V022636/1 and EP/Y028783/1. The authors wish to
 acknowledge the support of colleagues at Lancaster University and Shell.

\bibliographystyle{imsart-nameyear} 
\bibliography{Ref}       


\end{document}